# Electric and Magnetic Walls on Dielectric Interfaces

Changbiao Wang

*ShangGang Group, 70 Huntington Road, Apartment 11, New Haven, CT 06512, USA*

Sufficient conditions of the existence of electric or magnetic walls on dielectric interfaces are given for a multizone uniform dielectric waveguiding system. If one of two adjacent dielectric zones supports a TEM field distribution while the other supports a TM (TE) field distribution, then the common dielectric interface behaves as an electric (magnetic) wall, that is, the electric (magnetic) field line is perpendicular to the interface while the magnetic (electric) field line is parallel to the interface.

A better understanding of basic electromagnetic principles might be helpful in analyzing specific physical problems. In this Report, sufficient conditions of the existence of electric and magnetic walls on the dielectric interfaces are given for a mutizone uniform dielectric waveguiding system, which is intrinsically loss-free and translationally invariant in the guiding direction. We will show:

If one of two adjacent dielectric zones supports a TEM field distribution while the other supports a TM (TE) field distribution, then the common dielectric interface behaves as an electric (magnetic) wall, that is, the electric (magnetic) field line is perpendicular to the interface while the magnetic (electric) field line is parallel to the interface [1].

Suppose that the mutizone dielectric waveguiding system guides electromagnetic waves in the *z*-direction. The functions $\mathbf{E}(\mathbf{r})$ and $\mathbf{H}(\mathbf{r})$ are the electric and magnetic distributions in the *x-y* plane, the axial components satisfy the Helmholtz equation

$$\nabla_\perp^2 (E_z, H_z) + k_\perp^2 (E_z, H_z) = 0, \qquad (1)$$

and the transverse components are given by [2]

$$\mathbf{E}_\perp = -ik_\perp^{-2}\left(k_z \nabla_\perp E_z + \omega\mu \nabla_\perp H_z \times \hat{\mathbf{z}}\right), \qquad (2)$$

$$\mathbf{H}_\perp = -ik_\perp^{-2}\left(k_z \nabla_\perp H_z - \omega\varepsilon \nabla_\perp E_z \times \hat{\mathbf{z}}\right), \qquad (3)$$

where $\nabla_\perp$ is the transverse gradient operator, $k_\perp$ is the transverse wave number, defined by $k_\perp^2 = \omega^2 \mu\varepsilon - k_z^2$, $\omega$ is the frequency, $k_z$ is the axial wave number, and $\varepsilon$ and $\mu$ are, respectively, the permittivity and permeability, $\hat{\mathbf{z}}$ is the axial unit vector, *i* is the imaginary unit, and $\mathbf{E}(\mathbf{r})$ and $\mathbf{H}(\mathbf{r})$ satisfy all transverse electromagnetic boundary conditions. Consequently, the solution to the source-free guiding system can be written as $\{\mathbf{E}(\mathbf{r}), \mathbf{H}(\mathbf{r})\} \exp[i(\omega t - k_z z)]$.

Suppose that $\hat{\mathbf{n}}$ and $\hat{\mathbf{t}}$ are, respectively, the normal and tangential unit vectors on the dielectric boundary in the *x-y* plane, with $\hat{\mathbf{n}} = \hat{\mathbf{t}} \times \hat{\mathbf{z}}$, $\hat{\mathbf{t}} = \hat{\mathbf{z}} \times \hat{\mathbf{n}}$, and $\hat{\mathbf{z}} = \hat{\mathbf{n}} \times \hat{\mathbf{t}}$, as shown in Fig. 1. Dot-multiplying both sides of Eq. (2) by $\hat{\mathbf{t}}$ and Eq. (3) by $\hat{\mathbf{n}}$ respectively, we have

$$E_t = -ik_\perp^{-2}\left[k_z(\hat{\mathbf{t}} \cdot \nabla_\perp E_z) - \omega\mu(\hat{\mathbf{n}} \cdot \nabla_\perp H_z)\right], \qquad (4)$$

$$H_n = -ik_\perp^{-2}\left[k_z(\hat{\mathbf{n}} \cdot \nabla_\perp H_z) - \omega\varepsilon(\hat{\mathbf{t}} \cdot \nabla_\perp E_z)\right]. \qquad (5)$$

Figure 1 shows the cross section for the multizone uniform dielectric waveguiding system. Below we will show that, if zone (I) supports a TEM field distribution while zone (II) supports a TM field distribution, then the dielectric interface of zone (I) and zone (II) behaves as an electric wall, that is, the electric field line is perpendicular to the interface while the magnetic field line is parallel to the interface.

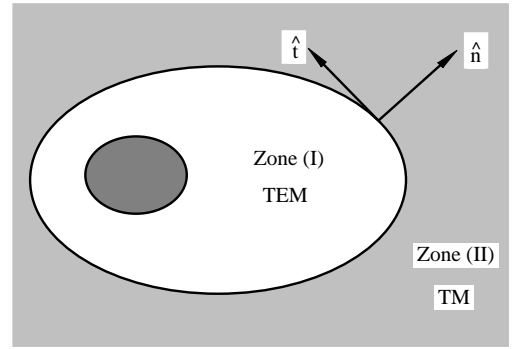

Fig. 1. Cross section for a multizone uniform dielectric waveguiding system. A TEM field distribution is supported in zone (I), where $E_z \equiv 0$ and $H_z \equiv 0$, while a TM field distribution is supported in zone (II), where $H_z \equiv 0$.

Since a TEM field structure is supported in zone (I), $(k_\perp^{(I)})^2 = \omega^2 \mu^{(I)} \varepsilon^{(I)} - k_z^2 = 0$ must hold in terms of Eqs. (2) and (3) to make the TEM fields physical. Accordingly, we have $(k_\perp^{(II)})^2 = \omega^2 \mu^{(II)} \varepsilon^{(II)} - k_z^2 \neq 0$ in zone (II) because the two zones have different dielectric parameters so that $\mu^{(I)} \varepsilon^{(I)} \neq \mu^{(II)} \varepsilon^{(II)}$. On the interface, from boundary conditions we have $E_z^{(I)} = E_z^{(II)} = 0$, leading to $\hat{\mathbf{t}} \cdot \nabla_\perp E_z^{(II)} = 0$. On the zone-(II) boundary, from Eqs. (4) and (5) we have

$$E_t^{(II)} = -i(k_\perp^{(II)})^{-2} \left[ k_z (\hat{\mathbf{t}} \cdot \nabla_\perp E_z^{(II)}) - \omega \mu^{(II)} (\hat{\mathbf{n}} \cdot \nabla_\perp H_z^{(II)}) \right] = 0, \quad (6)$$

$$H_n^{(II)} = -i(k_\perp^{(II)})^{-2} \left[ k_z (\hat{\mathbf{n}} \cdot \nabla_\perp H_z^{(II)}) - \omega \varepsilon^{(II)} (\hat{\mathbf{t}} \cdot \nabla_\perp E_z^{(II)}) \right] = 0. \quad (7)$$

where $\hat{\mathbf{n}} \cdot \nabla_\perp H_z^{(II)} = 0$ is employed due to $H_z \equiv 0$ in zone (II) for a TM field structure. From Eqs. (6) and (7) and boundary conditions we have $E_t^{(I)} = 0$ and $H_n^{(I)} = 0$ on zone-(I) boundary, that is, the common interface is an electric wall.

Similarly, we can show that if zone (I) supports a TEM field distribution while zone (II) supports a TE field distribution, then the common dielectric interface behaves as a magnetic wall.

As we have known, the TEM field structure requires that the transverse wave number $k_\perp = 0$ must hold. From this, we can draw another conclusion: Two dielectric zones, which have different dielectric parameter products ($\mu\varepsilon$), cannot support TEM field distributions at the same time, because of the failure of the simultaneous holding of $k_\perp = 0$. This result can be used to explain why a one-dielectric-filled microstrip transmission line or metallic coaxial waveguide can support a TEM mode while a multi-dielectric-filled one cannot support any TEM modes [3,4].

From Eqs. (2) and (3), we have

$$\mathbf{E}_\perp \cdot \mathbf{H}_\perp = k_\perp^{-2} (\nabla_\perp E_z) \cdot (\nabla_\perp H_z). \quad (8)$$

As we know, the electric and magnetic fields for a TEM field structure are perpendicular each other. From Eq. (8), we can see that the transverse components of the electric and magnetic fields for a TM or TE field structure are also perpendicular each other; however, for a hybrid EH field structure, they may not.